# Multiply Robust Estimator Circumvents Hyperparameter Tuning of Neural Network Models in Causal Inference

Mehdi Rostami[1,3] | Olli Saarela[2,3]

[1]Email: mehdi.rostamiforooshani@
mail.utoronto.com (**Correspondence**)

[2]Email: olli.saarela@utoronto.com

[3]Biostatistics, Dalla Lana School of Public
Health, University of Toronto, ON, Canada

**Abstract**

Estimation of the Average Treatment Effect (ATE) is often carried out in 2 steps, wherein the first step, the treatment and outcome are modeled, and in the second step the predictions are inserted into the ATE estimator. In the first steps, numerous models can be fit to the treatment and outcome, including using machine learning algorithms. However, it is a difficult task to choose among the hyperparameter sets which will result in the best causal effect estimation and inference. Multiply Robust (MR) estimator allows us to leverage all the first-step models in a single estimator. We show that MR estimator is $n^r$ consistent if one of the first-step treatment or outcome models is $n^r$ consistent. We also show that MR is the solution to a broad class of estimating equations, and is asymptotically normal if one of the treatment models is $\sqrt{n}$-consistent. The standard error of MR is also calculated which does not require a knowledge of the true models in the first step. Our simulations study supports the theoretical findings.

## 1 | INTRODUCTION

The estimation of the Average Treatment Effect (ATE) typically involves two steps. In the first step, a model is developed for the treatment and outcome. In the second step, predictions from the first step are used to estimate the ATE. There is a wealth of literature on various aspects of this process, including the types of estimators that can be used to calculate the causal effect [1, 2, 3, 4, 5], the models that are most appropriate to use in the first step [6, 7, 8, 9, 10], and the performance of causal effect estimators in cases where the models used in the first step converge slowly or are complex [11, 12].

There are several estimators, such as IPW [13], AIPW [3], and Robinson [14] that can be used to estimate the ATE. In the first step of the process, multiple statistical machine learning (ML) models can be trained with possibly many hyperparameter sets, while only a pair of models for the outcome and propensity score is needed. The choice of which models to use in the second step is often based on how well they fit the data in the first step, even though this does not necessarily result in an optimal estimation of the ATE.

The out-of-sample predictive measures such as the Area Under the Curve (AUC) (of the treatment model) and $R^2$ of the outcome model can be used. Rostami and Saarela [15] utilized the geometric average of AUC and $R^2$ as a criterion for selecting the best model. However, in causal inference, as the potential outcomes are not observed (are missing), and thus cannot be used in the training or selecting of the models, and thus the first step of model selection may not result in an optimal estimation of the causal effect. The super learner method [16] which could be used to construct ensemble predictions for the treatment and outcome avoids selecting a pair of models as the final weighted average predictions will be consistent if one of the input models is so. However, the same challenge still holds for utilizing super learner which is estimating the weighted average coefficients using the criterion that the predicted values should be as close as possible to the observed data (in the validation data set). Additionally, the model selection and ensemble prediction may be sensitive to the choice of the validation set used. If the validation set is not representative of the overall data distribution, they may perform poorly on out-of-sample data. Further, violation of the assumptions for using super learner such as independence of the data, and well-calibrated candidate algorithms [11], can impact the performance of the causal parameter estimator. These drawbacks of the super learner motivate utilizing an estimator that circumvents the need for hyperparameter tuning or selecting any models.

The MR estimator is a method for estimating ATE that was introduced in the field of missing data research [17]. It can make use of the models developed in the first step which would eliminate the need for hyperparameter tuning. The MR estimator is a consistent estimator of ATE if any of the models for the propensity score or outcome are also consistent and does need the knowledge of which one is consistent. It is an empirical maximum likelihood estimator that requires the solution of a constrained convex optimization problem. However, due to numerical issues with this method, an alternative optimization method was proposed by Han [18] that theoretically and numerically results in a multiply robust estimator of ATE. Wang [19] utilized MR for causal inference when the treatment is binary and Naik et al. [20] extended the MR to treatments with multiple categories. In both of these research, the models used in the first step are parametric models, and thus $\sqrt{n}$-consistent estimators. The current work is more similar to that of Wang [19] in the sense that we consider a binary treatment, but the major difference is that we are interested in using NN-based prediction models (or other non-parametric ML models) rather than parametric ones in the first-step, which are consistent but with a slower rate than $\sqrt{n}$. For the MR estimator to be a $\sqrt{n}$-consistent estimator, at least one of the treatment or outcome models must also be a $\sqrt{n}$-consistent estimator. This can be a stringent requirement in practice if ML models are used in the first step. The AIPW and nAIPW estimators have the property that if the selected PS and outcome models converge as quickly as $n^{\frac{1}{4}}$, the estimator is $\sqrt{n}$-consistent, under certain regulatory conditions and with the use of cross-fitting [12]. However, unlike in the AIPW and nAIPW estimators, for the MR estimator, we do not need to have the knowledge of which model is $\sqrt{n}$-consistent. Thus, there is a trade-off between the properties of MR and AIPW, and nAIPW estimators - one property is gained at the expense of another. In cases where it is unclear which model is consistent, MR estimator is preferable.

In terms of inference on the MR estimator, Wang [19] discussed and derived the asymptotic normality and variance estimator. One limitation of their proposed estimator is that it requires the knowledge of which treatment and/or outcome models are consistent, which goes against the purpose of using MR instead of AIPW. Therefore, the authors recommend using bootstrapping to estimate the standard error of the MR estimator.

The main goal of this paper is to compare the performance of the MR estimator to the normalized nAIPW estimator when the first-step models have a slow convergence rate (such as when neural networks are used, as in Farrell et al. [8]). The second objective is to develop an asymptotic variance estimator for MR that does not require knowledge of the consistent first-step models. The comparisons are performed for both low- and high-dimensional scenarios.

Through simulations, we will also use the same MR estimator when the first-step models are neural network models (which are slower to converge than parametric models, at a rate of around $n^{\frac{1}{4}}$ [8]). We will use simulations to study the performance of MR in terms of bias, variance, and root mean square error in the presence of instrumental variables (as defined in Angrist and Pischke [21]) and confounder variables in the data.

In this paper, we will study the impact of the number of trained models in the first step on the performance of the MR estimator, both with and without the use of oracle models for the propensity score and outcome. We will also compare MR to the normalized AIPW estimator [15, 22], using selection criteria based on $AUC$, $R^2$, and the geometric average of these prediction measures, *geo*. Additionally, we will provide a detailed mathematical derivation of the MR estimator and its multiple robustness property for slower rates than $\sqrt{n}$. In our theorems, we assume general functional forms for the outcome and treatment predictors, rather than parametric forms, and we prove $n^r$ consistency with $r \leq \frac{1}{2}$. We also study theoretically and numerically the proposed asymptotic variance estimator for the MR estimator. The proof for the asymptotic normality and the derivation of the asymptotic variance estimator will also be included.

This paper is organized as follows. In Section 2, we define the notation, specify the problem setting and the causal parameter to be estimated, and provide a brief overview of the normalized AIPW estimator. In Section 3, we review the MR estimator and outline its theoretical properties. In Section 4, we describe our simulation scenarios and present the results in Section 4.1. We conclude the paper in Section 6 with a discussion of the results and future work. The proofs of the lemmas and theorems are provided in the appendix 8.

## 2 | BACKGROUND

Let data $\mathbf{O} = (O_1, O_2, ..., O_n)$ be generated by a data generating process $P$, where $O_i$ is a finite-dimensional random vector $O_i = (Y_i, A_i, W_i)$, with $Y$ as the outcome, $A$ as the treatment and $\mathbf{W} = (X_c, X_y, X_{iv}, X_{irr})$ the covariates, where we assume $A = f_1(X_c, X_{iv})$, and $Y = f_2(A, X_c, X_y)$, for some random functions $f_1, f_2$. The covariates are partitioned so that $X_c$ is the set of confounders, $X_{iv}$ is the set of instrumental variables, $X_y$ is the set of y-predictors (independent of the treatment), and $X_{irr}$ is a set of given noise or irrelevant inputs (Figure 1). $P$ represents the true joint probably distribution of $\mathbf{O}$ and $\hat{P}_n$ is its sample version. Let $\hat{P}_n$ be any distribution of $(Y, A, W)$ such that the marginal distribution of $W$ is given by its empirical distribution, and the conditional distribution of $Y|(A = a, W)$ has a conditional mean equal to a given estimator $\mathbb{E}[Y|A = a, W]$. Let $Q^1$ represent the expected outcome of the treated group, where $Q^1 := Q(1, W) = \mathbb{E}[Y|A = 1, W]$, and $Q^0$ represent the expected outcome of the untreated group, where $Q^0 := Q(0, W) = \mathbb{E}[Y|A = 0, W]$. Also, let $g(W)$ be the propensity score, defined as $g(W) = \mathbb{E}[A|W]$. All expectations are taken with respect to $P$. The symbol ˆ on the population-level quantities indicates the corresponding finite sample estimator.

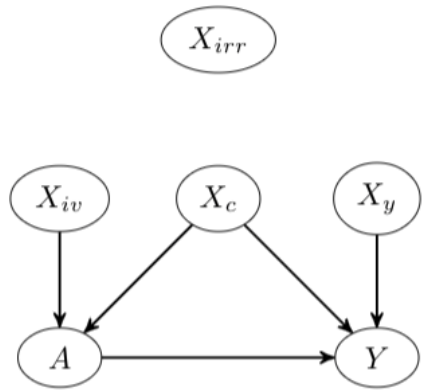

**Figure 1** The causal relationship between $A$ and $y$ in the presence of other factors in an observational setting.

### 2.1 | Problem Setup and Assumptions

The fundamental problem of causal inference is that individual-level causality cannot be identified since each person can experience only one value of $A$. Thus, it is customary to focus on estimating a population-level causal parameter, in this particular, the Average Treatment Effect (ATE),

$$\beta_{ATE} = \mathbb{E}[Y^1 - Y^0] = \mathbb{E}\left[\mathbb{E}[Y^1 - Y^0|W]\right]. \tag{1}$$

To identify the parameter, the following assumptions must be satisfied. The first assumption is conditional independence, ignorability, or unconfoundedness, which states that given the confounders, the potential outcomes are independent of the treatment assignments $(Y^0, Y^1 \perp A|W)$. The second assumption is positivity, which requires that the assignment of treatment groups is not deterministic $(0 < Pr(A = 1|W) < 1)$ (Van der Laan and Rose [11], page 344). The third assumption is consistency, which states that the observed outcomes are equal to their corresponding potential outcomes $(Y^A = Y)$. There are also other assumptions that are made, such as the order in which the covariates $W$, outcome and treatment are measured, and the assumption that subjects are independent and identically distributed (IID).

### 2.2 | Doubly Robust Estimator

A variety of estimators have been proposed for estimating the Average Treatment Effect (ATE), including Augmented Inverse Probability Weighting (AIPW) and Normalized, Augmented Inverse Probability Weighting (nAIPW). These are given by

$$\hat{\beta}_{AIPW} = \frac{1}{n} \sum_{i=1}^{n} \left( \frac{A_i(Y_i - \hat{q}_i^1)}{\hat{g}_i} - \frac{(1 - A_i)(Y_i - \hat{q}_i^0)}{1 - \hat{g}_i} \right) + \frac{1}{n} \sum_{i=1}^{n} \hat{q}_i^1 - \hat{q}_i^0, \text{ and}$$

$$\hat{\beta}_{nAIPW} = \sum_{i=1}^{n} \left( \frac{A_i(Y_i - \hat{q}_i^1)w_i^{(1)}}{\sum_{j=1}^{n} A_j w_j^{(1)}} - \frac{(1 - A_i)(Y_i - \hat{q}_i^0)w_i^{(0)}}{\sum_{j=1}^{n}(1 - A_j)w_j^{(0)}} \right) + \frac{1}{n} \sum_{i=1}^{n} \hat{q}_i^1 - \hat{q}_i^0, \tag{2}$$

where $\hat{q}_i^k = \hat{q}(k, W_i) = \hat{\mathbb{E}}[Y_i | A_i = k, W_i]$, and $\hat{g}_i = \hat{\mathbb{E}}[A_i | W_i]$. [15] demonstrated that nAIPW should be favored over AIPW in the datasets where strong confounders and instrumental variables exist and complex algorithms such as NNs are used in the first step.

In the second step of the estimation procedure, predictions of the treatment (i.e. propensity score) and the outcome $\hat{q}_i^k, k = 0, 1$, are inserted in the estimators (2). Generalized Linear Models (GLM), any relevant Machine Learning algorithm such as tree-based algorithms and their ensemble [23], SuperLearner [16], or Neural Network-based models (such as ours) can be applied as prediction models for the first step prediction task. We will use the Joint NN (jNN) and the Double NN (dNN), proposed by Rostami and Saarela [22]. jNN and dNN refer to NN architectures to estimate the outcome and treatment; jNN is one single NN with both treatment and outcome as the output nodes, and dNN refers to two separate NNs that predict outcome and treatment. We note that the theoretical properties of the AIPW and nAIPW are not guaranteed to hold if certain assumptions such as the assumption that first-step models are $\sqrt{n}$-consistent or assumptions such as those outlined by Chernozhukov et al. [12] of the first step models do not hold.

## 3 | MULTIPLE ROBUST ESTIMATOR

The multiply robust estimator of ATE (1) is

$$\hat{\beta} = \sum_{i=1}^{n} \hat{\beta}_i^1 - \hat{\beta}_i^0, \tag{3}$$

such that

$$\hat{\beta}_i^1 = \frac{A_i Y_i}{n_1 (1 + \hat{\gamma}^T \hat{C}_i^1)},$$
$$\hat{\beta}_i^0 = \frac{(1 - A_i) Y_i}{n_0 (1 + \hat{\rho}^T \hat{C}_i^0)}, \tag{4}$$

where $n_1$ and $n_0$ is the size of treatment and control groups, and

$$\hat{C}_i^1 = \begin{pmatrix} \hat{g}_i^1 - \hat{\mathbb{E}}\hat{g}^1 \\ ... \\ \hat{g}_i^K - \hat{\mathbb{E}}\hat{g}^K \\ \hat{Q}_i^1(1) - \hat{\mathbb{E}}\hat{Q}^1(1) \\ ... \\ \hat{Q}_i^L(1) - \hat{\mathbb{E}}\hat{Q}^L(1) \end{pmatrix}, \hat{C}_i^0 = \begin{pmatrix} \hat{\mathbb{E}}\hat{g}^1 - \hat{g}_i^1 \\ ... \\ \hat{\mathbb{E}}\hat{g}^K - \hat{g}_i^K \\ \hat{Q}_i^1(0) - \hat{\mathbb{E}}\hat{Q}^1(0) \\ ... \\ \hat{Q}_i^L(0) - \hat{\mathbb{E}}\hat{Q}^L(0) \end{pmatrix} \tag{5}$$

with $\hat{\mathbb{E}}\hat{g}^k$ and $\hat{\mathbb{E}}\hat{Q}^l(j)$ representing the average of the $\hat{g}^k$ and $\hat{Q}^l(j)$ over all observations ($i = 1, ..., n$), and $\gamma = (\gamma_1, ..., \gamma_{K+L})^T$ and $\rho = (\rho_1, ..., \rho_{K+L})^T$ are $K + L$ dimensional vectors that satisfy

$$\sum_{i=1}^{n} \frac{A_i \hat{C}_i^1}{1 + \hat{\gamma}^T \hat{C}_i^1} = 0, \quad \sum_{i=1}^{n} \frac{(1 - A_i) \hat{C}_i^0}{1 + \hat{\rho}^T \hat{C}_i^0} = 0, \tag{6}$$

with the propensity score and outcome estimators

$$\{\hat{g}^1, \hat{g}^2, ..., \hat{g}^K\}$$
$$\{\hat{Q}^1(j), \hat{Q}^2(j), ..., \hat{Q}^L(j)\}. \tag{7}$$

This way of defining the MR estimator maximizes the joint probability of $W, Y$, given $A = 1$.

**Lemma 1.** The estimator (4) maximizes the joint probability of $W, Y$, given $A = 1$.

**Lemma 2.** Given that one of the propensity score estimators is a consistent estimator of the true propensity score, the MR estimator is consistent.

**Lemma 3.** If one of the outcome regression models is correctly specified, then the MR estimator is consistent.

The proofs of these lemmas are left to the Appendix. The lemmas do not assume that we know which models are consistent, or consistent with a faster rate. The rate of consistency of MR depends on the fastest rate of the models used in the estimator. Now assuming that one of the above lemmas holds true, we have

**Theorem 3.1.** Given that either one of the propensity score estimators is a consistent estimator of the true propensity score or one of the outcome regression estimators of the outcome, the MR estimator is consistent.

*Proof.* By Lemmas 2 and 3. □

**Theorem 3.2.** MR estimator is a solution to an estimating equations system

$$\sum_{i=1}^{n} A_i u_i (Y_i - \beta^1) = 0,$$
$$\sum_{i=1}^{n} (1 - A_i) v_i (Y_i - \beta^0) = 0,$$

(8)

where $\beta = \beta^1 - \beta^0$, $u_i = f_{Y^1,W}(y_i, w_i)$, and $v_i = f_{Y^0,W}(y_i, w_i)$. Given that one of the outcome or PS models is $\sqrt{n}$-consistent, the MR estimator is $\sqrt{n}$-consistent and asymptotically normal, that is,

$$(\hat{\beta}_{MR} - \beta) \xrightarrow{d} N(0, \frac{\hat{\sigma}^2}{n}),$$

(9)

with

$$\hat{\sigma}^2 = n \sum_{i=1}^{n} \left( A_i \hat{u}_i^2 (Y_i - \beta_{MR}^1)^2 + (1 - A_i) \hat{v}_i^2 (Y_i - \beta_{MR}^0)^2 \right),$$

(10)

where

$$\hat{u}_i = \frac{1}{1 + \gamma^T C_i^1} / \left( \sum_{i=1}^{n} \frac{A_i}{(1 + \gamma^T C_i^1)} \right), \qquad \sum_{j=1}^{n} \frac{A_j C_j^{1,k}}{1 + \gamma^T C_j^1} = 0,$$
$$\hat{v}_i = \frac{1}{1 + \rho^T C_i^0} / \left( \sum_{i=1}^{n} \frac{1 - A_i}{(1 + \rho^T C_i^0)} \right), \qquad \sum_{j=1}^{n} \frac{(1 - A_j) C_j^{0,k}}{1 + \rho^T C_j^0} = 0,$$

(11)

for $k = 1, ..., K + L$.

*Remark 1.* The MR estimating equations (8) belong to a larger class of estimating equations

$$\sum_{i=1}^{n} A_i u_i (Y_i - \beta^1) - \eta_1 \left( 1 - A_i u_i \right) = 0,$$
$$\sum_{i=1}^{n} (1 - A_i) v_i (Y_i - \beta^0) + \eta_0 (1 - (1 - A_i) v_i) = 0,$$

(12)

where $\eta_1 = \eta_0 = 0$. This general class is similar to the general class of estimating equations that includes Inverse Probability Weighting (IPW), normalized Inverse Probability Weighting (nIPW), and AIPW estimators (Lunceford and Davidian [3], equation 10). Note that $u_i$ (and similarly $v_i$) act as the reciprocal of the propensity score. Further, the MR asymptotic variance estimator (51) matches with that of nIPW where $1/g_i$ is replaced by $u_i$ and $1/(1 - g_i)$ is replaced by $v_i$.

# 4 | SIMULATIONS

The objectives of the simulation study are 1) to demonstrate how the number of trained first-step models impacts the MR estimator, 2) to compare the performance of MR with nAIPW, 3) to visualize if the MR estimator is consistent (which is sufficient to show it's asymptotic unbiasedness), 4) to contrast the proposed asymptotic variance estimator with the Monte Carlo variance of the MR estimator, and 5) to visualize how the dimension of covariate space impacts the distribution of the MR estimator and its asymptotic variance estimator in a scenario where all the causal inference assumptions (Section 2.1) are satisfied.

In the simulation study, we generated 100 independent samples to compare the prediction methods jNN, and dNN by inserting their predictions in the nAIPW (causal) estimators (2), or utilizing them to calculate MR. For this purpose we allow the underlying relationship between the inputs and the output to be non-linear. We fixed the sample sizes to be $n = 750$ and $n = 7500$, with the number of covariates $p = 32$ and $p = 300$, respectively. We generate the confounding, instrumental variable, y-predictors, and noise/irrelevant inputs independently with the same sizes $\#X_c = \#X_{iv} = \#X_y = \#X_{irr} = 8, 75$ (summing to 32 and 300) from the Multivariate Normal (MVN) distribution as each set of covariates $X \sim \mathcal{N}(\mathbf{0}, \Sigma)$, with $\Sigma_{kj} = \rho^{j-k}$ and $\rho = 0.5$. Let $\beta = 1$. The models to generate the treatment assignment and outcome were specified as

$$
\begin{aligned}
A &\sim Ber\Big(\frac{1}{1 + e^{-\eta}}\Big), \text{ with } \eta = f_a(X_c)\gamma_c + g_a(X_{iv})\gamma_{iv}, \\
Y &= 3 + A + f_y(X_c)\gamma'_c + g_y(X_y)\gamma_y + \epsilon,
\end{aligned}
\tag{13}
$$

The functions $f_a, g_a, f_y, g_y$ select 30% of the columns and apply interactions and non-linear functions listed below (14). The strength of instrumental variable and confounding effects were chosen as $\gamma_c, \gamma'_c, \gamma_y \sim Unif(r_1, r_2)$ and $\gamma_{iv} \sim Unif(r_3, r_4)$ where $(r_1 = r_3 = 0)$, and $(r_2 = r_4 = 0.25)$. The non-linearities for each pair of covariates are randomly selected among the following functions:

$$
\begin{aligned}
l(x_1, x_2) &= e^{\frac{x_1 x_2}{2}} \\
l(x_1, x_2) &= \frac{x_1}{1 + e^{x_2}} \\
l(x_1, x_2) &= \Big(\frac{x_1 x_2}{10} + 2\Big)^3 \\
l(x_1, x_2) &= \big(x_1 + x_2 + 3\big)^2 \\
l(x_1, x_2) &= g(x_1) \times h(x_2)
\end{aligned}
\tag{14}
$$

where $g(x) = -2I(x \leq -1) - I(-1 \leq x \leq 0) + I(0 \leq x \leq 2) + 3I(x \geq 2)$, and $h(x) = -5I(x \leq 0) - 2I(0 \leq x \leq 1) + 3I(x \geq 1)$, or $g(x) = I(x \geq 0)$, and $h(x) = I(x \geq 1)$.

In order to avoid imbalanced treatment groups, the generated datasets in which the number of subjects in the treatment or control group is less than 25% were ignored and new ones were generated. Also, in order to examine the accuracy of the proposed asymptotic variance estimator, we have considered a low-confounding scenario with ($r_2 = r_4 = 0.01$) for two cases of low and high dimensional covariate space. In this side experiment, the only model included in the MR estimator is the Oracle model.

The networks' activation function is Rectified Linear Unit (ReLU), with 3 hidden layers as large as the input size (p), with $L_1$ regularization and batch size equal to $3 * p$ and 200 epochs. The Adaptive Moment Estimation (Adam) optimizer [24] with a learning rate 0.01 and momentum 0.95 were used to estimate the network's parameters. The total number of hyperparameter settings is 32: 2 scenarios for the two types of NN models (jNN and dNN), 2 NN depth variations ([q, p, q] and [p, p, p], where $q = p/4$), 2 $L_1$ regularization strengths (0.01, 0.1), 4 targetted $L_1$ regularization strengths (0, 0.1, 0.3, 0.7). These scenarios are exactly the same as the ones in the articles Rostami and Saarela [15] and Rostami and Saarela [22].

As in practice, to calculate AIPW/nAIPW, the true models are unknown and counterfactuals are unobserved, the prediction measures of the outcome and treatment should be used to choose the best model by the K-fold cross-validation, to insert in the estimators in step 2. $R^2$ and $AUC$ each provide insights into the outcome and treatment models, respectively, but in our framework, both models should be satisfactory. To measure the goodness of the prediction models (jNN and dNN) for causal inference purposes, we define and utilize a statistic which is a compromise (geometric average) between $R^2$ and $AUC$, here referred to as $geo$,

$$
geo(R, D) = \sqrt[3]{R^2 \times D \times (1 - D)},
\tag{15}
$$

where $D = 2(AUC - 0.5)$, the Somers' D index. The $geo$ measure was not utilized in the optimization process (i.e. training the neural networks or cross-validation), and is rather introduced here to observe if the compromise between $R^2$ and $AUC$ agrees with the models that capture more confounders than IVs. We will refer to $geo(R, D)$ simply as $geo$.

To compare the results with the truth in some of the results, we have used the true, or oracle, treatment, and outcome models Also, as we have 100 independent estimations for all the estimators and asymptotic variance, we can use bootstraping to estimate confidence intervals for them.

## 4.1 | Results

Figures 2 and 3 illustrate the performance of MR when the number of ML models trained in the first step increases by either including or excluding the single true model (the oracle model which is $\sqrt{n}$-consistent). Each dot is calculated by averaging 10 estimators using 10 models randomly selected among all the 32 models with different hyperparameter sets, outlined in the

**Figure 2** Comparison of MR estimator when different number of first step prediction models are fed into the estimator, and dataset is not large (n=750). In the first steps all NN models from jNN and dNN are used.

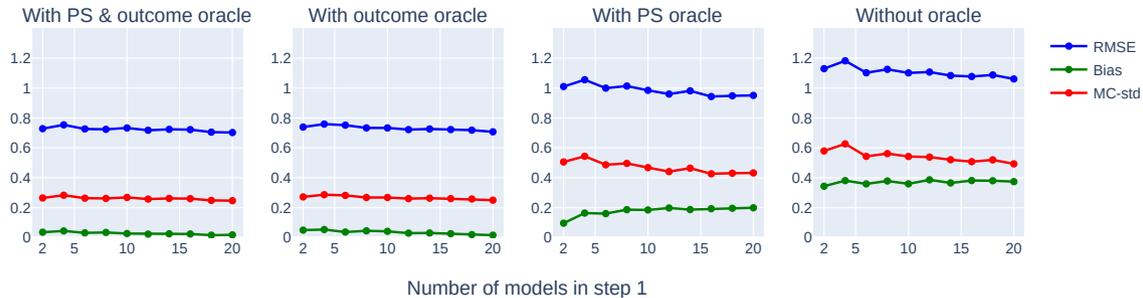

Number of models in step 1

previous section. It can be observed that for both small and large sample size settings, the performance (MR's bias, variance, or RMSE) does not significantly change irrespective of whether the predictions of the dNN, jNN, or their combination are used. As expected, a significant drop is observed by comparing the scenarios where the true outcome model is included in the first step models. When the true outcome model is not present but the true PS model is present, we observe a small improvement over the scenario where no oracle model is included in the first step models.

Figures 4 compares the performance of MR with the nAIPW which is a doubly robust estimator needing a choice of the best-performing first-step model. In the first 2 scenarios, the first-step prediction models (dNN or jNN) that minimize the RMSE are chosen; in the second two scenarios, the first-step prediction models (dNN or jNN) that maximize the $geo$ are chosen. The third and fourth pairs correspond to maximizing $R^2$ and $AUC$, respectively. The leftmost two settings are unrealistic, as we do not know the bias and variance of the estimator in real problems, but it can provide an ideal case with which other cases can be compared. When $AUC$ is used, neither jNN nor dNN shows favorable results in both small and large datasets. Also, it can be seen that no matter which of $R^2$, $AUC$, or $geo$ is used, it is better not to use the dNN predictions, especially for the small sample size case. On the other hand, predictions from jNN models when $R^2$ or $geo$ is used provide the best performance among the non-MR methods for both large and small sample sizes. Another observation is that MR shows an inferior performance as compared to nAIPW when jNN is used in the first model and the hyperparameters are chosen based on the best $geo$.

Comparing variations of MR and normalized AIPW estimation, it can be seen that MR is not superior to the normalized AIPW, when $R^2$ or $geo$ is used.

It is difficult to illustrate the consistency of an estimator through simulations. However, as asymptotic unbiasedness and asymptotic efficiency imply consistency [25], an illustration of the asymptotic unbiasedness and efficiency is equivalent to an illustration of consistency of the estimator. Figure 5 is a visualization of the asymptotic unbiasedness and asymptotic efficiency of the MR estimator with NN-type predictions. We note that the first step models are NNs which are consistent but slower than $n^{-\frac{1}{2}}$ rate, but the MR estimator provably and experimentally is consistent (even if the rate is slow.)

The proposed asymptotic standard error of the MR estimator is plotted against the MR estimator's Monte Carlo standard deviation in Figure 6. It can be seen that when the outcome oracle model is among the first step models in calculating the MR estimator, the proposed asymptotic standard error is accurate. However, when the Oracle model is not used in MR estimator, the proposed asymptotic standard error is severely biased in the case of the high dimensional case. This can be due to the curse of dimensionality or violation of causal inference assumptions. To examine which one is the cause of this bias, we have plotted (Figure 7) the sampling distribution of the MR estimator and its standard error when neither of the causal inference assumptions (Section 2.1) is violated, for a high and a low dimensional scenario. It can be seen that the variance estimator is still severely biased in the high-dimensional scenario. Similar results (figures not presented here) were seen for other estimators such as nIPW and nAIPW.

## 5  |  APPLICATION: FOOD INSECURITY AND BMI

The Canadian Community Health Survey (CCHS) is a survey that gathers data on various aspects of health, including health status, healthcare utilization, and health determinants, for the Canadian population in multiple cycles. The 2021 CCHS includes data on individuals aged 12 or older living in the ten Canadian provinces and the three territorial capitals. The survey excludes individuals living on reserves and other Aboriginal settlements in the provinces and some other subpopulations, which together make up less than 3% of the Canadian population aged 12 and over. In most cycles of the survey, data has been collected on subjects' general health, chronic conditions, smoking habits, and alcohol use. The 2021 cycle also includes questions on topics such as food security, home care, sedentary behavior, and depression, among others. In addition to health-related information, the survey also covers labor market activities, income, socio-demographic characteristics, and other characteristics of the respondents.

**Figure 3** Comparison of MR estimator when a different number of first step prediction models are fed into the estimator, and the dataset is large (n=7500). In the first steps all NN models from jNN and dNN are used.

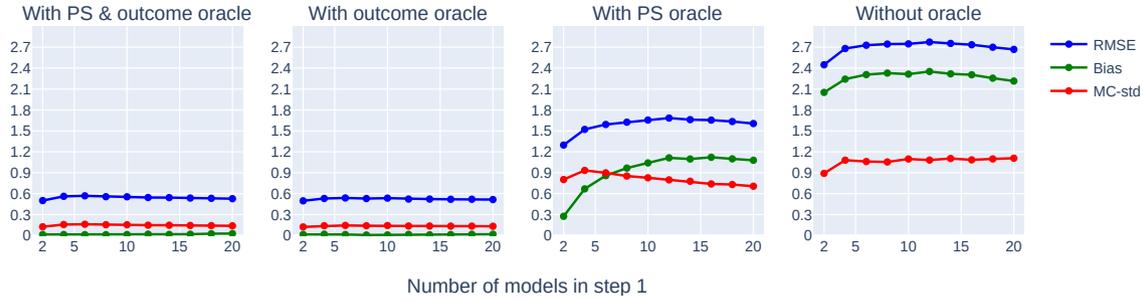

**Figure 4** Comparison of nAIPW and MR in terms of RMSE, bias, and standard error measurements where ($r_2 = r_4 = 0.25$). The three most right scenarios on the x-axis are MR estimators. The other scenarios on the x-axis are for the nAIPW where one of $geo$, $AUC$, and $R^2$ criteria have been used to choose the best first-step model among dNN or jNN models to insert into nAIPW.

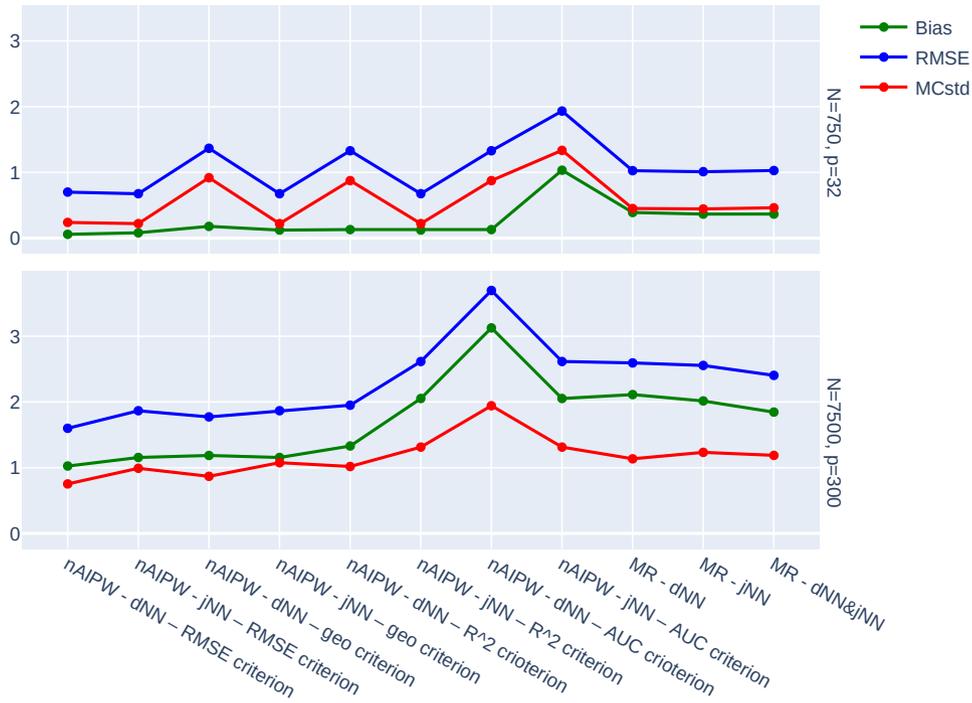

In this article, we use the CCHS dataset to study the causal relationship between food insecurity and body mass index (BMI). We also use other information collected in the CCHS that may be potential confounders, outcome predictors, or instrumental variables. The data for this survey needs to be analyzed using methods such as resampling or bootstrapping to estimate standard errors. However, in this case, we use the data to demonstrate how to use a jNN to analyze causal parameters in situations where the assumption of empirical positivity is violated. To reduce variability in the data, we focus on the sub-population of individuals aged 18-65.

**Figure 5** Consistency/asymptotic unbiasedness and asymptotic efficiency of MR. The predictions from jNN and dNN are used.

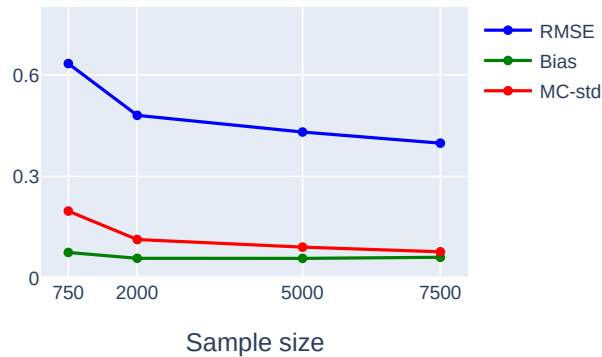

**Figure 6** Comparison of Monte Carlo standard deviation and asymptotic standard errors.

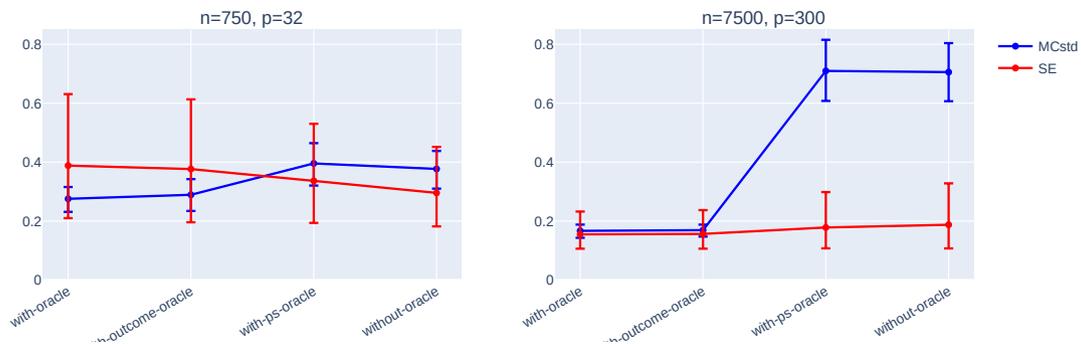

Figure 8 compares the ATE estimates and their 95% asymptotic confidence intervals using the nAIPW and MR methods. The nAIPW method is calculated for 8 different jNN hyperparameter scenarios, while the MR method uses a combination of these scenarios. The true ATE is unknown, but the MR method appears to be more stable than the nAIPW estimator across different hyperparameter settings. The confidence interval of the MR estimator is based on the proposed asymptotic variance estimator $\hat{\sigma}^2$ (51).

The code to implement the AIPW and MR methods in Python is available on the github page mentioned earlier. The dataset should include a continuous outcome, a binary treatment, and X-factors. The algorithms require predictions from multiple models for the triple $(Q(1), Q(0), g)$. Rows with missing values will be automatically removed.

# 6  |  DISCUSSION

The doubly robust estimators, such as AIPW, use one set of models for the outcome and propensity scores. If multiple models are developed in the first step, one set of models must be chosen (hard-selection) or an ensembling method such as the super learner (soft-selection) can be used to combine the predictions and insert them into the doubly robust estimator. Cross-validation is typically used to select the "best" outcome and propensity score models based on prediction performance metrics or to ensemble the predictions. In this article, we studied the MR estimator, which uses all trained models from the first step directly that does not require cross-validation to choose among the first-step models.

**Figure 7** The distribution of the MR estimator and the proposed asymptotic standard error for a low confounding scenario where $r_3 = r_4 = .01$. The figure shows that the curse of dimensionality greatly impacts the variance estimator, even if the causal inference assumptions (e.g. positivity assumption) are not violated.

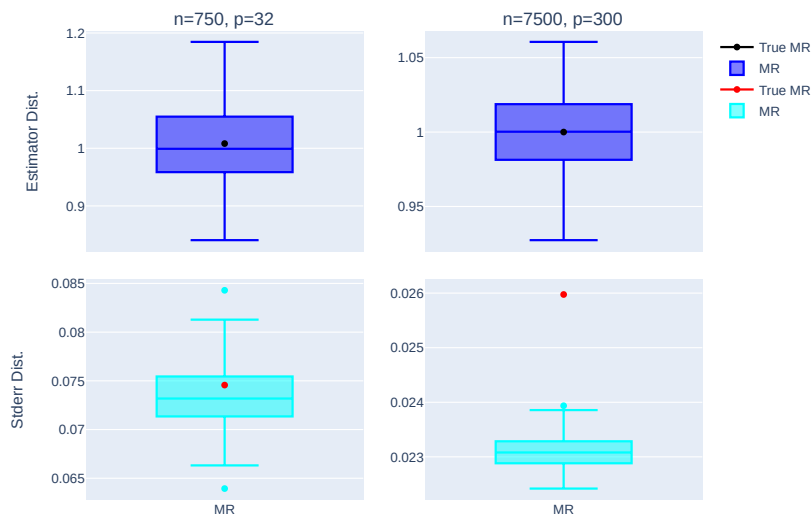

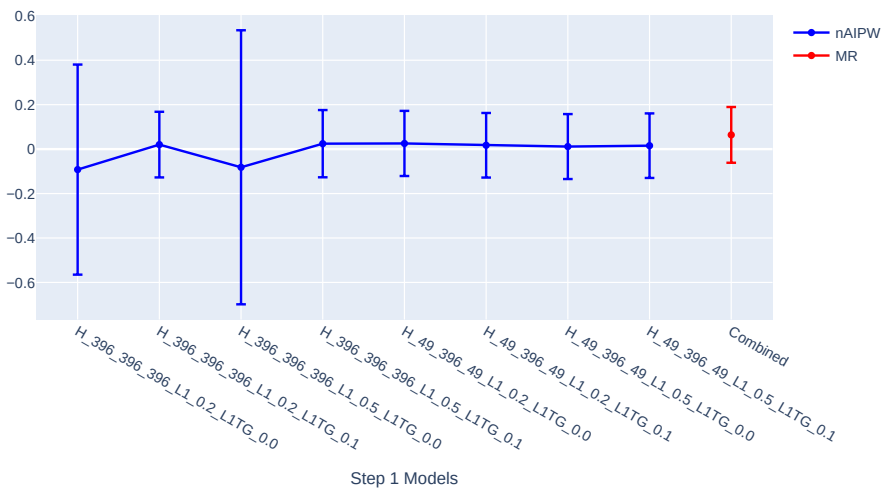

**Figure 8** The ATE estimates and their asymptotically calculated 95% confidence intervals with nAIPW and MR methods.

A new set of estimating equations (1) has been suggested, which bears resemblance to the one proposed in Lunceford and Davidian [3]. Both these classes of estimating equations allow for the generation of different parameters by setting two parameters $\eta_0$ and $\eta_1$. For example, by setting $\eta_0$ and $\eta_1$ to specific values, IPW, nIPW, and AIPW estimators can be produced, with nIPW being obtained when $\eta_0 = \eta_1 = 0$. Similarly, in our proposed class, MR can be obtained by setting $\eta_0 = \eta_1 = 0$. It is recommended that further investigation be conducted to explore other types of estimators that can be generated using our proposed class, which may perform better than AIPW and nAIPW.

The MR method has previously been used in missing data and causal inference studies where the first step models are parametric. In this article, we used Neural Network (NN) predictions to estimate MR and proved that the estimator is consistent if one of the outcome or propensity score models is consistent. We also investigated through simulations how the number of trained models in the first step affects the bias-variance trade-off of the MR estimator. We compared the MR method with the normalized AIPW (nAIPW) method and find that the results slightly favor the nAIPW method in terms of bias, MC variance, and RMSE.

It has been previously shown that if one of the trained propensity score models is $\sqrt{n}$-consistent, the MR estimator is asymptotically normal. For those proposed methods, however, model selection must be performed in the first step to calculate its asymptotic variance, which defeats the purpose of using MR without selecting a pair of trained models. We proposed an alternative variance estimator that does not require model selection. Theoretically, if the first step models are $\sqrt{n}$-consistent, the MR estimator is asymptotically normal with our proposed asymptotic variance estimator. The simulations showed that our proposed variance estimator performs satisfactorily for the low-dimensional case. For the high-dimensional case, the estimator does not perform well. Based on our research, it appears that the primary factor responsible for reducing the accuracy of the asymptotic variance estimator is dimensionality. We conjecture that the asymptotic variance estimator is affected by a bias up to a factor of $\sqrt{\frac{p}{n}}$, and further investigation and theoretical analysis are required to either support this claim or find an alternative solution. In addition, further research is required to investigate the appropriateness of our proposed asymptotic variance estimator in the case of first-step models that are slower than $\sqrt{n}$.

# 7 | PROGRAM CODES

The R-based python source code and examples for estimating the MR estimator and its asymptotic variance can be found on github.

## 8 | APPENDIX

**Lemma 4.** Let $g^k$'s and $Q^l$'s be the same as defined in (7), and $g = g(W)$ be the true propensity score. Then we have

$$\mathbb{E}\left[\frac{g^k - \mathbb{E}[g^k]}{g}\Big|A = 1\right] = 0, \quad k = 1, ..., K,$$

$$\mathbb{E}\left[\frac{Q^l(1) - \mathbb{E}[Q^l(1)]}{g}\Big|A = 1\right] = 0, \quad l = 1, ..., L,$$

$$\mathbb{E}\left[\frac{\mathbb{E}[g^k] - g^k}{1 - g}\Big|A = 0\right] = 0, \quad k = 1, ..., K,$$

$$\mathbb{E}\left[\frac{Q^l(0) - \mathbb{E}[Q^l(0)]}{1 - g}\Big|A = 0\right] = 0, \quad l = 1, ..., L.$$

(16)

where $Q^l(j) = Q^1(j, W)$.

*Proof.* We prove the first identity. By the positivity assumption, it is enough to show that

$$\mathbb{E}\left[\frac{g^k - \mathbb{E}[g^k]}{g}\Big|A = 1\right] P(A = 1) = 0 \tag{17}$$

By applying the total law of expectations twice, letting $I(.)$ be the indicator function, and using the definition of the propensity score function, we have

$$\mathbb{E}\left[\frac{g^k - \mathbb{E}[g^k]}{g}\Big|A = 1\right] P(A = 1) = \mathbb{E}\left[\frac{I(A = 1)\big(g^k - \mathbb{E}[g^k]\big)}{g}\Big|A = 1\right] P(A = 1) =$$

$$\mathbb{E}\left[\frac{I(A = 1)\big(g^k - \mathbb{E}[g^k]\big)}{g}\Big|A = 1\right] P(A = 1) +$$

$$\mathbb{E}\left[\frac{I(A = 1)\big(g^k - \mathbb{E}[g^k]\big)}{g}\Big|A = 0\right] P(A = 0) =$$

$$\mathbb{E}\left[\frac{I(A = 1)\big(g^k - \mathbb{E}[g^k]\big)}{g}\right] = \mathbb{E}\left[\mathbb{E}\left[\frac{I(A = 1)\big(g^k - \mathbb{E}[g^k]\big)}{g}\Big|X\right]\right] =$$

$$\mathbb{E}\left[\mathbb{E}\left[I(A = 1)\Big|X\right]\frac{g^k - \mathbb{E}[g^k]}{g}\right] = \mathbb{E}\left[g^k - \mathbb{E}[g^k]\right] = 0 \tag{18}$$

The second identity can be proven by following similar steps. The third identity can be proven by letting $D = A - 1$. Noted that in this theorem $g$ in the denumerators is assumed to be the true propensity score, but in later theorems, we do not need to know the true propensity scores. □

*Proof.* Proof of Lemma 1

Before proposing the estimator for ATE, consider the empirical joint probability distributions for the treatment and control groups $\{(y_i, w_i)|A_i = 1\}$ and $\{(y_i, w_i)|A_i = 0\}$. Define the probability functions $p_i = P(Y_i = y_i, W_i = w_i|A_i = 1)$, $q_i = P(Y_i = y_i, W_i = w_i|A_i = 0)$. Without any constraints on the $p_i$'s and $q_i$'s, it can be shown that $p_i = \frac{1}{n}$ and $q_i = \frac{1}{n}$ [26]. However, the sample version of (16) with the general form of the probability distribution functions $p_i$ and $q_i$ impose constraints on them

$$\sum_{i \in I_1} p_i \frac{g_i^k - \hat{\mathbb{E}}[g_i^k]}{g_i} = 0, \quad k = 1, ..., K,$$

$$\sum_{i \in I_1} p_i \frac{Q_i^l(1) - \hat{\mathbb{E}}[Q_i^l(1)]}{g_i} = 0, \quad l = 1, ..., L,$$

(19)

$$\sum_{i \in I_0} q_i \frac{\hat{\mathbb{E}}[g_i^k] - g_i^k}{1 - g_i} = 0, \quad k = 1, \dots, K,$$

$$\sum_{i \in I_0} q_i \frac{Q_i^l(0) - \hat{\mathbb{E}}[Q_i^l(0)]}{1 - g_i} = 0, \quad l = 1, \dots, L,$$

(20)

where $I_j = \{i : A_i = j\}$, for $j = 0, 1$, with cardinality $n_j = |I_j|$.

As

$$\frac{f_{Y^1, W | A=1}(y, w | A = 1)}{P(A = 1 | W = w)} P(A = 1) = \frac{f_{Y^1, W | A=1}(y, w | A = 1)}{P(A = 1 | W = w, Y^1 = y)} P(A = 1) = \frac{f_{Y^1, W | A=1}(y, w | A = 1)}{\frac{f_{Y^1, W | A=1}(y, w | A=1) P(A=1)}{f_{Y^1, W}(y, w)}} P(A = 1) = f_{Y^1, W}(y, w),$$

(21)

and $f_{Y^1, W | A=1}(y, w | A = 1) = f_{Y, W | A=1}(y, w | A = 1)$, we have

$$u_i = \frac{p_i \theta}{g(W_i)}.$$

(22)

Similarly,

$$v_i = \frac{q_i(1 - \theta)}{1 - g(W_i)}.$$

(23)

The empirical versions of $u_i$'s and $v_i$'s are obtained by replacing empirical versions of $p_i$'s and $q_i$'s in (22) and (23), respectively.

The goal is to estimate ATE

$$\beta = \mathbb{E}[Y^1] - \mathbb{E}[Y^0]$$

(24)

The MR estimator of ATE is

$$\hat{\beta} = \sum_{i=1}^n A_i \hat{u}_i Y_i - \sum_{i=1}^n (1 - A_i) \hat{v}_i Y_i.$$

(25)

This is motivated by the fact that $\mathbb{E}[Y^j] = \int y f_{Y^j}(y) dy = \int \int y f_{Y^j, W}(y, w) dw dy$, for $j = 0, 1$. For an observed sample, if the true $u$ and $v$ and potential outcomes are observed, the ideal (that is not available in practice) estimator is $\hat{\beta}_{ideal} = \sum_{i=1}^n u_i Y_i^1 - \sum_{i=1}^n v_i Y_i^0$; but the counterfactual outcomes are not observed in real life. The remedy is to calculate these sums for the observed outcomes and at the same time wight them by the reciprocal of the true propensity score in equations (22) and (23) to balance the distribution of the treatment and control groups due to the existence of the confounders, thus the estimator (25). Noted that in the process of calculating the MR estimator, true propensity scores are canceled and in reality, we do not need to know the true PS.

To estimate $u_i$ and $v_i$, we estimate $p_i$'s, and $q_i$'s (the latter is similar to the former, and thus we skip that). Here we assumed that $g$ in the denominator is one of the $K$ estimated propensity scores but we do not need to have the knowledge of which of them is the correctly specified propensity score model. Without the loss of generality, we can assume that $g^1$ is correctly specified, but for ease of notation, we still use $g$ in the denominator.

Empirical likelihood without constraints other than the normalization (summation of weights equals 1) is the same as the sample average for the estimation of expectations. However, the constraints (19) and (20) produce less straightforward estimators for the empirical probabilities (weights). Independence of subjects and the usage of the empirical likelihood method implies that in order to maximize the joint probability of $W, Y$, given $A = 1$, we must maximize the quantity

$$\max_{p_i : i \in I_1} \prod_{i \in I_1} p_i.$$

(26)

This results in a convex optimization problem with equality and inquality constraints:

$$\max_{p_i : i \in I_1} \prod_{i \in I_1} p_i,$$

subject to

$$p_i > 0, \quad i \in I_1,$$

$$\sum_{i \in I_1} p_i = 1,$$

$$\sum_{i \in I_1} p_i \frac{g_i^k - \hat{\mathbb{E}}[g_i^k]}{g_i} = 0, \quad k = 1, ..., K \tag{27}$$

$$\sum_{i \in I_1} p_i \frac{Q_i^l(1) - \hat{\mathbb{E}}[Q_i^l(1)]}{g_i} = 0, \quad l = 1, ..., L.$$

Equivalently, we can find $\max_{p_i : i \in I_1} \sum_{i \in I_1} log(p_i)$, with the same constraints as above. Applying the Lagrange multiplier, a solution to (27) is a solution to the following convex problem

$$\max_{p_i : i \in I_1} \sum_{i \in I_1} log(p_i) - \lambda^T \left\{ \sum_{i \in I_1} p_i \frac{C_i^j}{g_i} \right\}_{k=1}^{K+L} - \eta^T P - c(\sum_{i \in I_1} p_i - 1), \tag{28}$$

subject to

$$\frac{1}{\hat{p}_i} - \lambda^T C_i^1 / g_i - \eta_i - c = 0,$$

$$\eta_i p_i = 0, \ i \in I_1,$$

$$\eta_i \geq 0, \ i \in I_1, \tag{29}$$

$$\sum_{i \in I_1} p_i = 1,$$

$$\sum_{i \in I_1} p_i \frac{C_i^{1,k}}{g_i} = 0, \quad k = 1, ..., K + L.$$

where $\lambda$ is a $K + L$ dimensional parameter vector, and $\eta$ is a $|I_1|$-dimensional vector, and $P$ is the row vector of $p_i$, for $i \in I_1$ ].

By the first three conditions, we have that $\eta_i = 0, i \in I_1$, and

$$\hat{p}_i = \frac{1}{c + \lambda^T C_i^1 / g_i}. \tag{30}$$

By the forth and fifth conditions, we have

$$\sum_{i \in I_1} \frac{1}{c + \lambda^T C_i^1 / g_i} = 1,$$

$$\sum_{i \in I_1} \frac{C_i^{1,k} / g_i}{c + \lambda^T C_i^1 / g_i} = 0, \quad k = 1, ..., K + L. \tag{31}$$

By multiplying $\lambda_j$ to each identity to both sides of the second identity above and summing them, we have

$$\sum_{i \in I_1} \frac{\alpha_i}{c + \alpha_i} = \sum_{i \in I_1} 1 - \frac{c}{c + \alpha_i} = 0, \text{where } \alpha_i = \lambda^T C_i^1 / g_i. \tag{32}$$

Thus, by the first identity in (31), we have $c = n_1$. Thus

$$\hat{p}_i = \frac{1}{n_1} \frac{1}{1 + \tau^T C_i^1 / g_i}, \text{ where } \sum_{i \in I_1} \frac{C_i^{1,k} / g_i}{1 + \tau^T C_i^1 / g_i} = 0, \quad k = 1, ..., K + L, \tag{33}$$

where $\tau = \lambda / n_1$.

Without loss of generality, assuming that $\hat{g}^1$ is the correctly specified propensity score estimator, by substituting $\tau_1 = -1 + \gamma_1 \theta$, and $\tau_k = \gamma_k \theta$, for $k = 2, ..., K + L$ into (33), we can re-write $u_i$ as

$$\hat{u}_i = \theta \times \frac{1}{\theta n_1} \frac{1}{1 + \gamma^T C_i^1}, \quad \text{where} \sum_{i \in I_1} \frac{C_i^{1,k}}{1 + \gamma^T C_i^1} = 0, \quad k = 1, ..., K + L, \tag{34}$$

which does not depend on the true propensity score or the correctly specified estimator of the propensity score. Hence,

$$\hat{u}_i = \frac{1}{n_1} \frac{1}{1 + \gamma^T C_i^1}, \quad \text{where} \; A_i = 1, \; \sum_{j \in I_1} \frac{C_j^{1,k}}{1 + \gamma^T C_j^1} = 0, \quad k = 1, ..., K + L, \tag{35}$$

and

$$\hat{v}_i = \frac{1}{n_0} \frac{1}{1 + \rho^T C_i^0}, \quad \text{where} \; A_i = 0, \; \sum_{j \in I_0} \frac{C_j^{0,k}}{1 + \rho^T C_j^0} = 0, \quad k = 1, ..., K + L, \tag{36}$$

where $n_1$ and $n_0$ are the sizes of treatment and control groups. Or we can normalize $u_i$'s and $v_i$'s:

$$\hat{u}_i = \frac{1}{1 + \gamma^T C_i^1} / \left( \sum_{j=1}^n \frac{1}{1 + \gamma^T C_j^1} \right), \quad \text{where} \; A_i = 1, \; \sum_{j \in I_1} \frac{C_j^{1,k}}{1 + \gamma^T C_j^1} = 0, \quad k = 1, ..., K + L, \tag{37}$$

and

$$\hat{v}_i = \frac{1}{1 + \rho^T C_i^0} / \left( \sum_{j \in I_0} \frac{1}{1 + \rho^T C_j^0} \right), \quad \text{where} \; A_i = 0, \; \sum_{j \in I_0} \frac{C_j^{0,k}}{1 + \rho^T C_j^0} = 0, \quad k = 1, ..., K + L. \tag{38}$$

$$\square$$

**Lemma 5.** *If one of the propensity score estimators is correctly specified, then $\hat{\lambda}$ converges to zero in probability, that is $\hat{\lambda} \xrightarrow{p} 0$.*

*Proof.* We have that for all $j \in \{1, 2, ..., K + L\}$,

$$f(\vec{g}, \vec{Q}, \vec{\lambda}) = \sum_{i \in I_1} \frac{C_i^{1,j} / g_i}{1 + \lambda^T C_i^1 / g_i} = \begin{pmatrix} \sum_{i \in I_1} \frac{\hat{g}_i^1 - \hat{\mathbb{E}} \hat{g}^1}{g_i + \lambda^T C_i^1} \\ ... \\ \sum_{i \in I_1} \frac{\hat{g}_i^K - \hat{\mathbb{E}} \hat{g}^K}{g_i + \lambda^T C_i^1} \\ \sum_{i \in I_1} \frac{\hat{Q}_i^1(1) - \hat{\mathbb{E}} \hat{Q}^1(1)}{g_i + \lambda^T C_i^1} \\ ... \\ \sum_{i \in I_1} \frac{\hat{Q}_i^L(1) - \hat{\mathbb{E}} \hat{Q}^L(1)}{g_i + \lambda^T C_i^1} \end{pmatrix} = 0, \tag{39}$$

where $C_i^1$ contains $K$ estimators of the propensity score, and $L$ estimators of the outcome regression (for the treated group.) Without loss of generality, assume the correctly specified model is $g^1$, the first one, and for the ease of notation, we use the notation $g$ instead of $g^1$. And assume the true values for the propensity scores and outcome regressions are $g^{1,*}, ..., g^{K,*}, Q^{1,*}, ..., Q^{L,*}$, or in vector form, $\vec{g}^*, \vec{Q}^*$.

The first-order Taylor expansion of one of the entries in the left hand side of (39) around $(\vec{g}^*, \vec{Q}^*, \vec{0})$ is:

$$f(\vec{g}^*, \vec{Q}^*, \vec{0}) + \partial f_{\vec{\lambda}}(\vec{g}^*, \vec{Q}^*, \vec{0}) \vec{\lambda} + \partial f_{g^1}(\vec{g}^*, \vec{Q}^*, \vec{0})(g^1 - g^{1,*}) + \sum_{j=2}^K \partial f_{g^j}(\vec{g}^*, \vec{Q}^*, \vec{0})(g^j - g^{j,*}) +$$

$$\sum_{j=1}^L \partial f_{Q^j}(\vec{g}^*, \vec{Q}^*, \vec{0})(Q^j - Q^{j,*}), \tag{40}$$

where $\partial f_{\vec{\lambda}}(\vec{g}^*, \vec{Q}^*, \vec{0}) \vec{\lambda}$ could also be written as $\sum_{l=1}^{K+L} \partial f_{\lambda_l}(\vec{g}^*, \vec{Q}^*, \vec{0}) \lambda_l$.

Applying (40) on all observations in (39)

$$0 = n^{r-1} \sum_{i=1}^{n} A_i f(\vec{g}_i, \vec{Q}_i, \vec{\lambda}_i) \approx n^{r-1} \sum_{i=1}^{n} A_i f(\vec{g}_i^*, \vec{Q}_i^*, \vec{0}) + \left[\frac{1}{n} \sum_{i=1}^{n} A_i \partial f_{\vec{\lambda}}(\vec{g}_i^*, \vec{Q}_i^*, \vec{0})\right] n^r \hat{\vec{\lambda}} +$$

$$\left[\frac{1}{n} \sum_{i=1}^{n} A_i \partial f_{g_i^1}(\vec{g}_i^*, \vec{Q}_i^*, \vec{0})\right] n^r (\hat{g}_i^1 - g_i^{1,*}) + \sum_{j=2}^{K} \left[\frac{1}{n} \sum_{i=1}^{n} A_i \partial f_{g_i^j}(\vec{g}_i^*, \vec{Q}_i^*, \vec{0})\right] n^r (\hat{g}_i^j - g_i^{j,*}) +$$

$$\sum_{j=1}^{L} \left[\frac{1}{n} \sum_{i=1}^{n} A_i \partial f_{Q_i^j}(\vec{g}_i^*, \vec{Q}_i^*, \vec{0})\right] n^r (\hat{Q}_i^j - Q_i^{j,*}) + o_p(1) =$$

$$n^{r-1} \sum_{i=1}^{n} \frac{A_i}{g_i^{1,*}} \begin{pmatrix} g_i^{1,*} - \mathbb{E}\, g^{1,*} \\ ... \\ g_i^{K,*} - \mathbb{E}\, g^{K,*} \\ Q_i^{1,*}(1) - \mathbb{E}\, Q^{1,*}(1) \\ ... \\ Q_i^{L,*}(1) - \mathbb{E}\, Q^{L,*}(1) \end{pmatrix} - \frac{1}{n} \sum_{i=1}^{n} \begin{pmatrix} A_i \left(\frac{g_i^{1,*} - \mathbb{E}\, g^{1,*}}{g_i^{1,*}}\right)^2 \\ ... \\ A_i \left(\frac{g_i^{K,*} - \mathbb{E}\, g^{K,*}}{g_i^{1,*}}\right)^2 \\ A_i \left(\frac{Q_i^{1,*}(1) - \mathbb{E}\, Q^{1,*}(1)}{g_i^{1,*}}\right)^2 \\ ... \\ A_i \left(\frac{Q_i^{L,*}(1) - \mathbb{E}\, Q^{L,*}(1)}{g_i^{1,*}}\right)^2 \end{pmatrix} n^r \hat{\vec{\lambda}} +$$

$$\frac{1}{n} \sum_{i=1}^{n} \begin{pmatrix} A_i \frac{\mathbb{E}\, g^{1,*}}{(g_i^{1,*})^2} \\ -A_i \frac{g_i^{2,*} - \mathbb{E}\, g^{2,*}}{(g_i^{1,*})^2} \\ ... \\ -A_i \frac{g_i^{K,*} - \mathbb{E}\, g^{K,*}}{(g_i^{1,*})^2} \\ -A_i \frac{Q_i^{1,*}(1) - \mathbb{E}\, Q^{1,*}(1)}{(g_i^{1,*})^2} \\ ... \\ -A_i \frac{Q_i^{L,*}(1) - \mathbb{E}\, Q^{L,*}(1)}{(g_i^{1,*})^2} \end{pmatrix} n^r (\hat{g}_i^1 - g_i^{1,*}) +$$

$$\sum_{j=1}^{L} \frac{1}{n} \sum_{i=1}^{n} \begin{pmatrix} 0 \\ ... \\ \frac{A_i}{g_i^{1,*}} \\ ... \\ 0 \\ 0 \\ ... \\ 0 \end{pmatrix} n^r (\hat{g}_i^j - g_i^{j,*}) + \sum_{j=1}^{L} \frac{1}{n} \sum_{i=1}^{n} \begin{pmatrix} 0 \\ ... \\ 0 \\ \frac{A_i}{g_i^{1,*}} \\ 0 \\ ... \\ 0 \end{pmatrix} n^r (\hat{Q}(1)_i^j - Q(1)_i^{j,*}) + o_p(1). \tag{41}$$

The third term in (41) can be re-written as

$$\frac{1}{n} \sum_{i=1}^{n} \begin{pmatrix} \frac{A_i}{g_i^{1,*}} - A_i \frac{g_i^{1,*} - \mathbb{E}\, g^{1,*}}{(g_i^{1,*})^2} \\ -A_i \frac{g_i^{2,*} - \mathbb{E}\, g^{2,*}}{(g_i^{1,*})^2} \\ ... \\ -A_i \frac{g_i^{K,*} - \mathbb{E}\, g^{K,*}}{(g_i^{1,*})^2} \\ -A_i \frac{Q_i^{1,*}(1) - \mathbb{E}\, Q^{1,*}(1)}{(g_i^{1,*})^2} \\ ... \\ \frac{1}{n} \sum_{i=1}^{n} -A_i \frac{Q_i^{L,*}(1) - \mathbb{E}\, Q^{L,*}(1)}{(g_i^{1,*})^2} \end{pmatrix} n^r (\hat{g}_i^1 - g_i^{1,*}).$$

Thus, the summation of last three terms in (41) is

$$\frac{1}{n}\sum_{i=1}^{n}\begin{pmatrix} -A_i\frac{g_i^{1,*}-\mathbb{E}\,g^{1,*}}{(g_i^{1,*})^2} \\ -A_i\frac{g_i^{2,*}-\mathbb{E}\,g^{2,*}}{(g_i^{1,*})^2} \\ \cdots \\ -A_i\frac{g_i^{K,*}-\mathbb{E}\,g^{K,*}}{(g_i^{1,*})^2} \\ -A_i\frac{Q_i^{1,*}(1)-\mathbb{E}\,Q^{1,*}(1)}{(g_i^{1,*})^2} \\ \cdots \\ -A_i\frac{Q_i^{L,*}(1)-\mathbb{E}\,Q^{L,*}(1)}{(g_i^{1,*})^2} \end{pmatrix} n^r(\hat{g}_i^1-g_i^{1,*}) + n^{r-1}\sum_{i=1}^{n}\frac{A_i}{g_i^{1,*}}\begin{pmatrix} \hat{g}_i^1-g_i^{1,*} \\ \hat{g}_i^2-g_i^{2,*} \\ \cdots \\ \hat{g}_i^L-g_i^{L,*} \\ \hat{Q}(1)_i^1-Q(1)_i^{1,*} \\ \cdots \\ \hat{Q}(1)_i^K-Q(1)_i^{K,*} \end{pmatrix}.$$

Using the above simplified terms, the equation (41) can be further simplied to

$$n^{r-1}\sum_{i=1}^{n}\begin{pmatrix} \frac{A_i(\hat{g}_i^1-\mathbb{E}\,g^{1,*})}{g_i^{1,*}} \\ \cdots \\ \frac{A_i(\hat{g}_i^K-\mathbb{E}\,g^{K,*})}{g_i^{1,*}} \\ \frac{A_i(\hat{Q}_i^1(1)-\mathbb{E}\,Q^{1,*}(1))}{g_i^{1,*}} \\ \cdots \\ \frac{A_i(\hat{Q}_i^L(1)-\mathbb{E}\,Q^{L,*}(1))}{g_i^{1,*}} \end{pmatrix} + \frac{1}{n}\sum_{i=1}^{n}\begin{pmatrix} -A_i\left(\frac{g_i^{1,*}-\mathbb{E}\,g^{1,*}}{g_i^{1,*}}\right)^2 \\ \cdots \\ -A_i\left(\frac{g_i^{K,*}-\mathbb{E}\,g^{K,*}}{g_i^{1,*}}\right)^2 \\ -A_i\left(\frac{Q_i^{1,*}(1)-\mathbb{E}\,Q^{1,*}(1)}{g_i^{1,*}}\right)^2 \\ \cdots \\ -A_i\left(\frac{Q_i^{L,*}(1)-\mathbb{E}\,Q^{L,*}(1)}{g_i^{1,*}}\right)^2 \end{pmatrix} n^r\hat{\hat{\lambda}}-$$

$$n^{r-1}\sum_{i=1}^{n}\begin{pmatrix} A_i\frac{g_i^{1,*}-\mathbb{E}\,g^{1,*}}{(g_i^{1,*})^2} \\ A_i\frac{g_i^{2,*}-\mathbb{E}\,g^{2,*}}{(g_i^{1,*})^2} \\ \cdots \\ A_i\frac{g_i^{K,*}-\mathbb{E}\,g^{K,*}}{(g_i^{1,*})^2} \\ A_i\frac{Q_i^{1,*}(1)-\mathbb{E}\,Q^{1,*}(1)}{(g_i^{1,*})^2} \\ \cdots \\ A_i\frac{Q_i^{L,*}(1)-\mathbb{E}\,Q^{L,*}(1)}{(g_i^{1,*})^2} \end{pmatrix}(\hat{g}_i^1-g_i^{1,*}) + o_p(1).$$

Thus

$$n^{r-1}\sum_{i=1}^{n}\frac{A_i\hat{C}_i^{1,*}}{g_i^{1,*}} + Mn^r\hat{\hat{\lambda}} - Vn^{r-1}(\hat{g}_i^1-g_i^{1,*}) + o_p(1) = 0,$$

which implies

$$n^r\hat{\hat{\lambda}} = M^{-1}\left(-n^{r-1}\sum_{i=1}^{n}\frac{A_i\hat{C}_i^{1,*}}{g_i^{1,*}} + Vn^{r-1}(\hat{g}_i^1-g_i^{1,*})\right) + o_p(1),$$

which means $\hat{\hat{\lambda}}$ is $n^r$-consistent, if one of the propensity score models is consistent with a faster than or equal rate to $n^r$. Noted that, by (16) equations, $\sum_{i=1}^{n}\frac{A_i\hat{C}_i^{1,*}}{g_i^{1,*}}$ converges to zero in probability.

In above equations, we have used these derivative:

$$\frac{\partial}{\partial \lambda_j} \frac{C_k^1}{g^1 + \vec{\lambda}^T C^1} = -\frac{C_j^1 C_k^1}{(g^1 + \vec{\lambda}^T C^1)^2},$$

$$\frac{\partial}{\partial g^1} \frac{C_1^1}{g^1 + \vec{\lambda}^T C^1} = \frac{g^1 + \lambda^T C^1 - C_1^1(1 + \lambda_1)}{(g^1 + \vec{\lambda}^T C^1)^2},$$

$$\frac{\partial}{\partial g^1} \frac{C_k^1}{g^1 + \vec{\lambda}^T C^1} = -\frac{C_1^1(1 + \lambda_1)}{(g^1 + \vec{\lambda}^T C^1)^2}, \tag{42}$$

$$\frac{\partial}{\partial g^j} \frac{C_k^1}{g^1 + \vec{\lambda}^T C^1} = -\frac{C_k^1 \lambda_j}{(g^1 + \vec{\lambda}^T C^1)^2},$$

$$\frac{\partial}{\partial Q^j} \frac{C_k^1}{g^1 + \vec{\lambda}^T C^1} = -\frac{C_k^1 \lambda_j}{(g^1 + \vec{\lambda}^T C^1)^2}.$$

□

*Proof.* Proof of Lemma 2.

Lemma 5 proves that given one of the propensity scores is consistent faster than $n^r$ rate, we have than $\hat{\gamma} \xrightarrow{p} 0$ (and similarly it can be proven that $\hat{\rho} \xrightarrow{p} 0$) with $n^r$ rate. As $u_i = \frac{\theta p_i}{g_i}$ and $\hat{p}_i = \frac{1}{n_1} \frac{2}{1 + \hat{\lambda}^T C_i^1 / g_i}$ (22), we have:

$$\hat{\beta}^1 = \frac{1}{n_1} \sum_{i=1}^n \frac{A_i Y_i}{g_i + \hat{\lambda}^T C_i^1} \xrightarrow{p} \mathbb{E} \frac{AY}{g}. \tag{43}$$

Similarly, $\hat{\beta}^0 \xrightarrow{p} E \frac{(1-A)Y}{1-g}$. The same is true for the normalized version

$$\hat{\beta}^1 = \sum_{i=1}^n \frac{A_i Y_i}{g_i + \hat{\lambda}^T C_i^1} / \left( \sum_{i=1}^n \frac{A_i}{g_i + \hat{\lambda}^T C_i^1} \right) \xrightarrow{p} \mathbb{E} \frac{AY}{g} / \mathbb{E} \frac{A}{g}, \tag{44}$$

□

*Proof.* Proof of Lemma 3:

Let one of the outcome regression models be consistent, say, $\hat{Q}_j(1)^1$, where the super script 1 refers to the first model trained among $L$ first step models, with no loss of generality. The constraints in (6) contains

$$\sum_{i=1}^n \left( u_i A_i \hat{Q}_i^1(1) - \hat{\mathbb{E}} \hat{Q}^1(1) \right) = 0, \quad \sum_{i=1}^n \left( v_i(1 - A_i)\hat{Q}_i^1(0) - \hat{\mathbb{E}} \hat{Q}^1(0) \right) = 0, \tag{45}$$

which can be re-written as

$$\sum_{i=1}^n A_i \hat{u}_i \hat{Q}_i(1)^1 = \frac{1}{n} \sum_{i=1}^n \hat{Q}_i(1)^1,$$

$$\sum_{i=1}^n (1 - A_i) \hat{v}_i \hat{Q}_i(0)^1 = \frac{1}{n} \sum_{i=1}^n \hat{Q}_i(0)^1. \tag{46}$$

Thus,

$$\hat{\beta}^1 = \sum_{i=1}^n A_i \hat{u}_i Y_i = \sum_{i=1}^n A_i \hat{u}_i (Y_i - \hat{Q}_i(1)^1) + \frac{1}{n} \sum_{i=1}^n \hat{Q}_i(1)^1 \xrightarrow{p} \mathbb{E} \left[ A u^* (Y - Q(1)^{1,*}) \right] + \beta^1 = \beta^1, \tag{47}$$

where $\hat{u}$ converges to $u^*$. Noted that the last equality holds as $A(Y - Q(1)) = A(Y^1 - Q(1)) = 0$. Similarly, $\sum_{i=1}^n (1 - A_i)\hat{v}_i Y_i \xrightarrow{p} \beta^0$, which proves the consistency of the MR estimator. Thus, by having the consistency of one of the outcome models, the MR estimator $\sum_{i=1}^n A_i \hat{u}_i Y_i$ is consistent. □

*Proof.* Proof of theorem 3.2:

MR estimator is a solution to an estimating equations system

$$\sum_{i=1}^{n} \phi_i^1 = \sum_{i=1}^{n} A_i u_i (Y_i - \beta^1) = 0,$$

$$\sum_{i=1}^{n} \phi_i^0 = \sum_{i=1}^{n} (1 - A_i) v_i (Y_i - \beta^0) = 0,$$

(48)

where $\phi = (\phi^1, \phi^0)^T$, $\beta = \beta^1 - \beta^0$, $u_i = P(Y_i = y_i, W_i = w_i | A_i = 1)$, and $v_i = P(Y_i = y_i, W_i = w_i | A_i = 0)$. By the asymptotic results of the M-estimators [27]

$$\sqrt{n} \begin{pmatrix} \hat{\beta}^1 \\ \hat{\beta}^0 \end{pmatrix} \xrightarrow{d} MVN \left( \begin{pmatrix} \beta^1 \\ \beta^0 \end{pmatrix}, V \right)$$

where $V = \mathbf{I}^{-1}(O)\mathbf{B}(O)\mathbf{I}^{-1}(O)^T$ is the sandwich estimator with

$$\mathbf{I}(O) = \frac{-1}{n} \sum_{i=1}^{n} \dot{\phi}_i = \frac{1}{n} \sum_{i=1}^{n} \begin{pmatrix} A_i u_i & 0 \\ 0 & (1 - A_i) v_i \end{pmatrix} = \frac{1}{n} \begin{pmatrix} 1 & 0 \\ 0 & 1 \end{pmatrix},$$

and

$$\mathbf{B}(O) = \frac{1}{n} \sum_{i=1}^{n} \phi_i \phi_i^T = \frac{1}{n} \sum_{i=1}^{n} \begin{pmatrix} A_i u_i^2 (Y_i - \beta^1)^2 & 0 \\ 0 & (1 - A_i) v_i^2 (Y_i - \beta^0)^2 \end{pmatrix}.$$

Thus the sandwich estimator of the variance-covariance matrix of $\begin{pmatrix} \hat{\beta}^1 \\ \hat{\beta}^0 \end{pmatrix}$ is

$$V = \begin{pmatrix} n \sum_{i=1}^{n} A_i u_i^2 (Y_i - \beta^1)^2 & 0 \\ 0 & n \sum_{i=1}^{n} (1 - A_i) v_i^2 (Y_i - \beta^0)^2 \end{pmatrix}.$$

This implies that

$$\sqrt{n}(\hat{\beta} - \beta) \xrightarrow{d} N(0, \sigma^2),$$

(49)

with

$$\sigma^2 = n \sum_{i=1}^{n} \left( A_i u_i^2 (Y_i - \beta^1)^2 + (1 - A_i) v_i^2 (Y_i - \beta^0)^2 \right),$$

(50)

or

$$var(\hat{\beta}_{MR}) = \sum_{i=1}^{n} \left( A_i u_i^2 (Y_i - \beta^1)^2 + (1 - A_i) v_i^2 (Y_i - \beta^0)^2 \right)$$

(51)

To estimate $\sigma^2$ we replace the estimated and $\beta^{j}$'s with their MR estimators and $u_i, v_i$ with

$$\hat{u}_i = \frac{1}{1 + \gamma^T C_i^1} / \left( \sum_{i=1}^{n} \frac{A_i}{(1 + \gamma^T C_i^1)} \right), \qquad \sum_{j=1}^{n} \frac{A_j C_j^{1,k}}{1 + \gamma^T C_j^1} = 0,$$

$$\hat{v}_i = \frac{1}{1 + \rho^T C_i^0} / \left( \sum_{i=1}^{n} \frac{1 - A_i}{(1 + \rho^T C_i^0)} \right), \qquad \sum_{j=1}^{n} \frac{(1 - A_j) C_j^{0,k}}{1 + \rho^T C_j^0} = 0,$$

(52)

for $k = 1, ..., K + L$.

$\square$

$\square$